# Prediction magnetocrystalline anisotripy Fe-Rh thin films via machine leaning

Eun Sung Jekal [1,*], Hyunwoo Park[1,]

1 Department of Materials, ETH Zurich, Zurich 8093, Switzerland

* Correspondence: everjekal@gamil.com, Tel.: +41 76 510 63 96

**Abstract**

Least absolute shrinkage and selection operator (Lasso) was originally formulated for least squares models and this simple case reveals a substantial amount about the behavior of the estimator. It also shows that the coefficient estimates need not be unique if covariates are collinear. Using this Lasso technique, we analyze a magnetocrystalline anisotropy energy which is a long-standing issue in transition-metal thin films, expectially for Fe-Rh thin film systems on a MgO substrate. Our LASSO regression took advantage of the data obtained from first principles calculations for single slabs with seven atomic-layers of binary Fe-Rh films on MgO(001). In the case of Fe-Rh thin films, we have successfully found a linear behavior between the MCA energy and the anisotropy of orbital moments.

**Keywords**: machine learning, magnetism, magnetocrystalline anisotropy

## 1. Introduction

### 1.1. Magnetocystalline anisotropy

After Moore's Law broke few years ago, researchers have started to look for alternatives materials[1-5]. One of a promising candidate way is magnetic random access memory (MRAM) due its advantages such as the non-volatility, high speed process as well as low power consumption[6-16]. Beside, in another perspective, it is necessary to develop themagnetic tunnel junction element which consists of insulating and magnetic metal thin films with strongly preferred perpendicular magnetization with respect to the film plane as Fig. 1.

In this respect, materials containing strong spin orbit coupling (SOC), for example $4d$ and $5d$ elements can be expected to have perpendicular magnetocrystalline anisotropy (MCA)[17-19]. This has been confirmed by calculations, for instance, in some Fe-Rh binary systems[20-22]. In such systems, while the magnetic moments are carried out by the Fe atoms or other 3d transition metal (TM) elements, the strong SOC is introduced by the non-magnetic such as Rh which is 5d TM elements[20]. However, it was shown theoretically that the MCA of thin films depends strongly not only on the choice of atomic elements, but also on the detail of the atomic layer alignment. Designing novel magnetic metal multilayer films with a strong perpendicular MCA is, therefore, a very demanding task. The difficulty is in part related to the fact that MCA is related to many non-trivial but competing factors such as the SOC strength and crystal field, in a way that the full understanding of the origin of MCA has been one of the greatest issues in the field of magnetism.

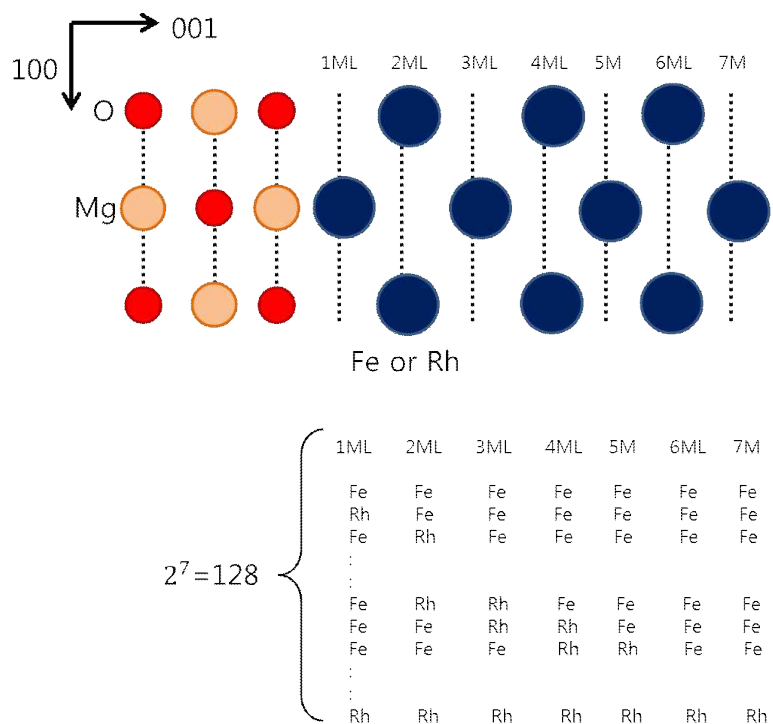

**Figure 1.** Calculation model and notation for atomic layer configurations of Fe-Rh/MgO. The perpendicular and in-plane directions are [001] and [100], respectively. All atomic layer patterns ($2^7$ = 128) expressed by binary configurations are considered as illustrated in the right side of the figure

## 1.2. Least absolute shrinkage and selection operator (Lasso)

Lasso was introduced in order to improve the prediction accuracy and interpretability of regression models by altering the model fitting process to select only a subset of the provided covariates for use in the final model rather than using all of them[23,24]. It was developed independently in geophysics, based on prior work that used the L1 penalty for both fitting and penalization of the coefficients, and by the statistician, Robert Tibshirani based on Breiman's non-negative garrote. Prior to lasso, themost widely usedmethod for choosing which covariates to include was stepwise selection, which only improves prediction accuracy in certain cases, such as when only a few covariates have a strong relationship with the outcome. However, in other cases, it can make prediction error worse. Also, at the time, ridge regression was the most popular technique for improving prediction accuracy. Ridge regression improves prediction error by shrinking large regression coefficients in order to reduce overfitting, but it does not perform covariate selection and therefore does not help to make the model more interpretable.

Lasso is able to achieve both of these goals by forcing the sum of the absolute value of the regression coefficients to be less than a fixed value, which forces certain coefficients to be set to zero, effectively choosing a simpler model that does not include those coefficients. This idea is similar to ridge regression, in which the sum of the squares of the coefficients is forced to be less than a fixed value, though in the case of ridge regression, this only shrinks the size of the coefficients, it does not set any of them to zero.

### 1.3. Link between MCA and Lasso

In an effort to understand the MCA of Fe-Rh films on a MgO substrate, using the LASSO technique, we examine the MCA and the anisotropy of orbital moments obtained from first principles calculations of Fe and Rh atoms to analyze the Bruno relation. Our analysis indeed confirms that the relation is applicable for systems.

### 2. crystal structure and method

Self-consistent Density Functional Theory calculations were performed based on the generalized gradient approximation (GGA) without SOC, by using the Full-potential Linearized Augmented Plane-Wave (FLAPW) method and Quantum Espresso. In all calculations, two-dimensional film geometries with additional vacuum regions have been considered. We model single slabs with seven-atomic layers of binary Rh-Fe film on top of three layers of MgO (001) substrates. All atomic-layer configurations, which include as many as $2^7=128$ possibilities for binary system, were considered in the calculations (see Fig. 1). A planewave cut-off |k + G| of 3.9 a.u. has been chosen, and we choose suitable muffin-tin radii of 2.4 bohr for Rh and 2.2 bohr for Fe. The MCA energy, $E_{MCA}$, is defined in our calculations as the energy eigenvalue difference between the magnetizations oriented along the in-plane [100] and perpendicular [001] directions with respect to the film plane. To determine the $E_{MCA}$, the second variational method for treating SOC by using the calculated eigenvectors in the SRA has been employed, and then the MCA energy has been calculated by using the force theorem.

The number of special k-points in the two-dimensional Brillouin zone (BZ) was 3600, and it is found that such k-point mesh was large enough to suppress numerical fluctuations in the MCA energy by less than 0.01 meV, which is a sufficient accuracy for the purpose of the present work. In regression analysis, we consider the linear regression model in which the MCA energy is expressed by using the orbital magnetic moment,

$$E_{MCA} = \sum_{i-1}^{6} A^{Co,i} M_{anis}^{Co,i} + B^{Fe,i} M_{anis}^{Fe,i} + D^{Fe,i} M_{anis}^{Fe,i}$$

Here, the $E_{MCA}$ has been expanded by using the descriptors, which for the present work, we choose the anisotropies and averages of orbital moments. The orbital moment anisotropy is defined as the difference between the orbital moments of the perpendicular and in-plane magnetizations for Fe or Rh species at the $i$th layer. Likewise, the orbital moment average is defined as the average of the perpendicular and inplane orbital moments of the respective atomic species. To eliminate the bias of choosing the descriptor, we consider that is not only the anisotropy, but also the average included. A leave-oneout cross validation, where the data is separated to 63 training data and one test data, was applied. Coefficients A, B, C and D in the Eq. 1 show how the descriptors are correlated to the parameters. For example, a large AAu,1 indicates that the anisotropy of orbital moment of the Au atomic layer placed at the top layer, i.e., layer 1, strongly affects the prediction of $E_{MCA}$. The A, B, C, and D coefficients are determined by using the training data, and then used to predict $E_{MCA}$ of the test data.

## 3. Results

The calculated results of MCA energy and anisotropy of orbital moments as functions of the atomic composition for each binary system are shown in Fig. 2.

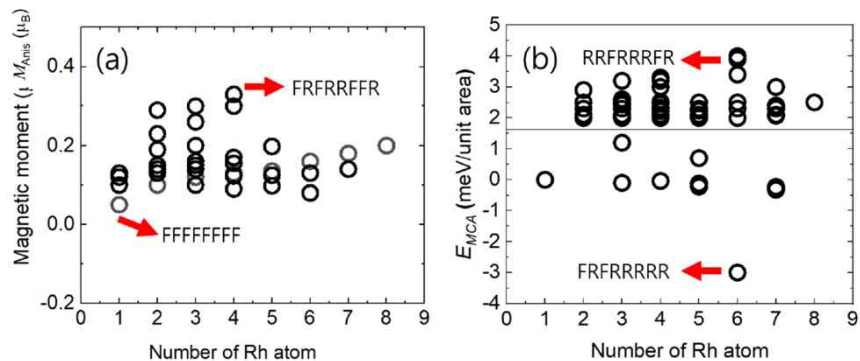

**Figure 1.** Results of anisotropy of orbital moments, $M_{anis}$, (a) and MCA energies, $E_{MCA}$, (b) as functions of the composition for Fe-Rh thin film systems. Atomic layer configurations are represented, e.g., by FRFRRFFR for FeRhFeRhRhFeFeRh/MgO, where F and R indicate Fe and Rh

atomic layers, respectively.

## 3. Results

The calculated results of MCA energy and anisotropy of orbital moments as functions of the atomic composition for each binary system are shown in Fig. 2. Both the computed values diverge largely even within the same atomic composition. For a specific atomic composition, e.g., for a Fe2Rh6 in the Fe-Rh binary system, the MCA energy varies largely from in-plane anisotropy with an $E_{MCA}$ of about 4.0 meV/unit-area, to a perpendicular one with an $E_{MCA}$ of about 1.5 meV/unit-area. Figure 3a shows a scatter plot between $E_{MCA}$ and $M_{anis}$ which suggest a linear relationship, as visualized with the red line. Hence, large values, implying the significance of all anisotropy of orbital moments. Figure 3c further shows the comparison between the calculated $E_{MCA}$ from first principles and the estimated ones using LASSO. These results show that the $E_{MCA}$ values predicted using LASSO agree well those obtained exlicitly by using first principles calculations. In the Fe-Rh binary systems, the EMCA varies largely from 4.2 meV/unit-area with a perpendicular anisotropy to -3.0 meV/unit-area with an inplane one. Fig. 2b plots between MCA energy, $E_{MCA}$, and anisotropy of orbital moments, $M_{anis}$, for Fe-Rh binary system. (b) calculated coefficients for anisotropy and average values of orbital moments in Eq. 1, where $M_{ani}$ and $M_{ave}$ indicate, respectively, the anisotropies and averages of the orbital moment. The numbers in the horizontal axis denote the atomic layer positions in the film from the surface to the interface. Comparison of the MCA energies calculated by the first principle calculations and estimated by using the LASSO coefficients. Red and blue solid circles show the data of training and test. Then I calculated by first principle calculations and those predicted using the LASSO coefficients are not satisfying. Indeed, the induced magnetic moments in Rh due to Fe. Combined with a strong SOC, such small splitting will lead to important spin-flip term to the SOC energy, resulting 111 in a breakdown of the Bruno relation. Additionally, in the present analysis, we considered only the orbital moment anisotropy and averages as the data descriptors for the MCA energy. The discrepancy between the first-principles calculated and the predicted $E_{MCA}$s reveals that such descriptors might not be sufficient to predict the MCA energy of systems with strong SOC.

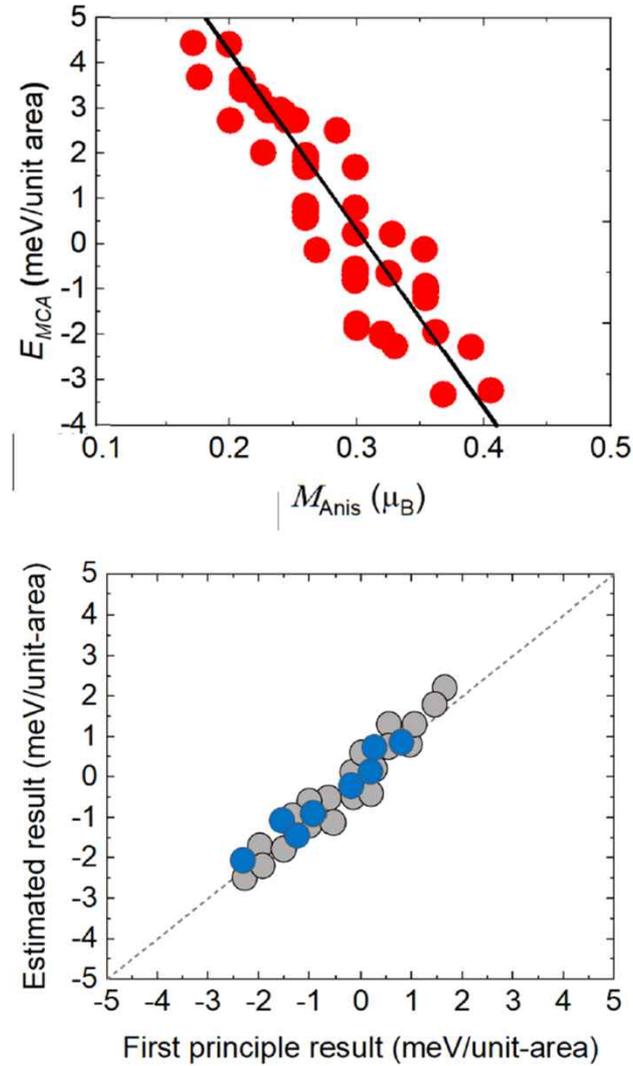

## 4. Conclusions

We found calculation results which are combined first-principles calculations (density functional theory) and LASSO which is machine learning analysis. Finally we show that the $E_{MCA}$ of the Fe-Rh thin films is strongly correlated to the anisotropy of orbital moments. Furthermore, the estimated $E_{MCA}$ shows a reasonable agreement with the calculated one.